\documentclass[twocolumn,aps,pre,superscriptaddress,10pt]{revtex4-2}

\usepackage{graphicx}
\usepackage[dvipsnames]{xcolor}
\usepackage{amsmath}
\usepackage{amssymb}
\usepackage{upgreek}
\usepackage[colorlinks=true,urlcolor=RoyalBlue,citecolor=RoyalBlue,linkcolor=Black]{hyperref}

\graphicspath{{.}}

\usepackage{xr-hyper}
\renewcommand{\figurename}{Figure}
\makeatletter
\renewcommand{\fnum@figure}{\small{\textbf{\figurename~\thefigure}}}
\renewcommand*{\@caption@fignum@sep}{ \small{$\mathbf{|}$} }
\makeatother

\renewcommand{\tablename}{Table}
\makeatletter
\renewcommand{\fnum@table}{\small{\textbf{\tablename~\thetable}}}
\makeatother

\begin{document}

\title{Deciphering the dual chemotaxis strategy of bacteria in porous media}

\author{S\"onke Beier}
\affiliation{Institute of Physics and Astronomy, University of Potsdam, D-14476 Potsdam, Germany}

\author{Veronika Pfeifer}
\affiliation{Institute of Physics and Astronomy, University of Potsdam, D-14476 Potsdam, Germany}

\author{Agniva Datta}
\affiliation{Institute of Physics and Astronomy, University of Potsdam, D-14476 Potsdam, Germany}

\author{Robert Gro{\ss}mann}
\affiliation{Institute of Physics and Astronomy, University of Potsdam, D-14476 Potsdam, Germany}

\author{Carsten Beta}
\email[Corresponding author:~]{beta@uni-potsdam.de}
\affiliation{Institute of Physics and Astronomy, University of Potsdam, D-14476 Potsdam, Germany}
\affiliation{Nano Life Science Institute (WPI-NanoLSI), Kanazawa University, Kakuma-machi, Kanazawa 920-1192, Japan}

\date{\today}

\begin{abstract}	
Chemotaxis of bacterial swimmers that move in a run-and-turn pattern is well studied in uniform bulk fluid.
It is primarily based on modulating the run time in dependence on the swimming direction with respect to the source of chemoattractant~(run time bias). 
Here, we provide evidence that the lophotrichously flagellated soil bacterium \textit{Pseudomonas putida} may also perform chemotaxis in porous media where the free path length is severely restricted.
Besides the classical run time bias, we identify a second chemotactic strategy:~the change in swimming direction upon a turn event is adjusted, so that the direction of the next run phase is biased towards the source of chemoattractant~(turn angle bias).
Agent based simulations, based on the experimentally observed statistical properties of the swimming pattern, indicate that turn angle bias is the predominant chemotaxis strategy of bacteria in porous environments. 
\end{abstract}

\maketitle

\noindent
\large{\textbf{INTRODUCTION}}

\vspace{0.3em}

\normalsize
\normalfont 
\noindent
Bacteria sense their environmental conditions and may respond to chemical stimuli by chemotaxis, the directed movement toward or away from a chemical source. 
Despite its high metabolic and energetic cost, about half of the known bacterial species can perform chemotactic motion~\cite{Wuichet2010,Sanchis-Lopez2021}. 
It provides significant benefits as the directed motion along a chemical gradient allows bacteria to access nutrients and to guide culture expansion towards more favorable conditions~\cite{Colin2021,Matilla2023}. 

Typically, bacteria swim in straight runs interrupted by short reorientation events.
To bias their direction of motion, bacteria modulate the frequency of the reorientation events leading to longer runs when swimming towards a source of chemoattractant~\cite{Berg1972}.
This is known as the run time bias.
In complex and confined habitats including tissues and sediments, swimming motility and chemotaxis are strongly affected by the geometry of the environment.
Nevertheless, also in these surroundings, directed movement is a key prerequisite for many essential processes such as tissue infection~\cite{Josenhans2002} or symbiosis with plants in the rhizosphere~\cite{Raina2019}.
In confined environments, however, runs are restricted by the mean free path length, so that bacteria frequently collide with the surrounding matrix or become trapped~\cite{Bhattacharjee2019b,Pfeifer2022}. 
If the mean free path length matches the average run length, bacteria reorient frequently enough to escape from traps but runs are, at the same time, long enough to enable efficient spatial exploration~\cite{Wolfe1989a, Raatz2015, Kurzthaler2021, datta2024intermittentrunmotilitybacteria}.
In very dense environments, in contrast, run lengths are severely reduced; in consequence, a chemotaxis strategy that is based on modulating the run length becomes inefficient~\cite{Bhattacharjee2021}.

An alternative mechanism to achieve a chemotactic drift is the turn angle bias: cells actively modulate the changes in swimming direction that are induced by turn events, depending on the gradient direction.
To navigate towards a source of chemoattractant, cells swimming away from the source will, on average, perform larger turns than cell swimming towards it.
For \textit{Escherichia coli}, a turn angle bias has been reported for chemotaxis in bulk fluid~\cite{Vladimirov2010,Saragosti2011,Pohl2017,Seyrich2018} and it was speculated to play a role during chemotaxis in porous environments~\cite{Bhattacharjee2021}.

To elucidate the relevance of a turn angle bias for chemotaxis in disordered confined surroundings, we focus on the soil bacterium \textit{Pseudomonas putida}.
The natural habitat of \textit{P.~putida} is a crowded, densely packed granular soil environment, where \textit{P.~putida} colonizes plant roots and shows chemotaxis towards root exudates~\cite{Lopez-Farfan2019}.
\textit{P.~putida} thus appears as a natural candidate to address the question of chemotactic navigation under confinement.
In aqueous bulk solution, the lophotrichously flagellated \textit{P.~putida} cells exhibit a complex multi-mode swimming pattern composed of three different run types:~the flagellar bundle may push, pull, or wrap around the cell body~\cite{Theves2013,Hintsche2017}.
Run phases are interrupted by stops and reversals in the swimming direction, resulting in small and large turning angles, respectively~\cite{Theves2013}.
In the presence of a chemoattractant gradient, the swimming pattern in uniform liquid surroundings exhibits a directional drift induced by a run time bias in wrapped swimming mode~\cite{Alirezaeizanjani2020}.
To address the question of chemotaxis strategies of bacteria in porous gels, we focus in this study on the chemotactic motion of \textit{P.~putida} in disordered matrices composed of different concentrations of agar.

In a classical swimming agar assay, bacteria are injected into semisolid agar and spread within several hours radially around the injection point.
This can be described as a diffusive spreading process based on random swimming motility, cell growth and division~\cite{Wolfe1989a}.
Chemotaxis can additionally promote the spreading of a culture~\cite{Cremer2019}.
For \textit{P.~putida}, it was shown that the swimming motility in agar can be divided into run phases of straight displacements and different turning events during which the bacterium stops and reorients.
In particular, in addition to the actively triggered short stops and directional reversals, known from motility in bulk fluid, longer lasting trapping events were observed in agar, which originate from mechanical arrest in the porous network~\cite{Pfeifer2022, datta2024intermittentrunmotilitybacteria}.
In this work, we characterize the chemotaxis strategy of \textit{P.~putida} in an agar network.
We show that chemotactic navigation of \textit{P.~putida} in porous matrices relies on both, run time bias and turn angle bias. \\

\noindent
\large{\textbf{RESULTS}}

\vspace{0.3em}

\begin{figure}[t]
\centering
\includegraphics[width=0.98\columnwidth, draft=false]{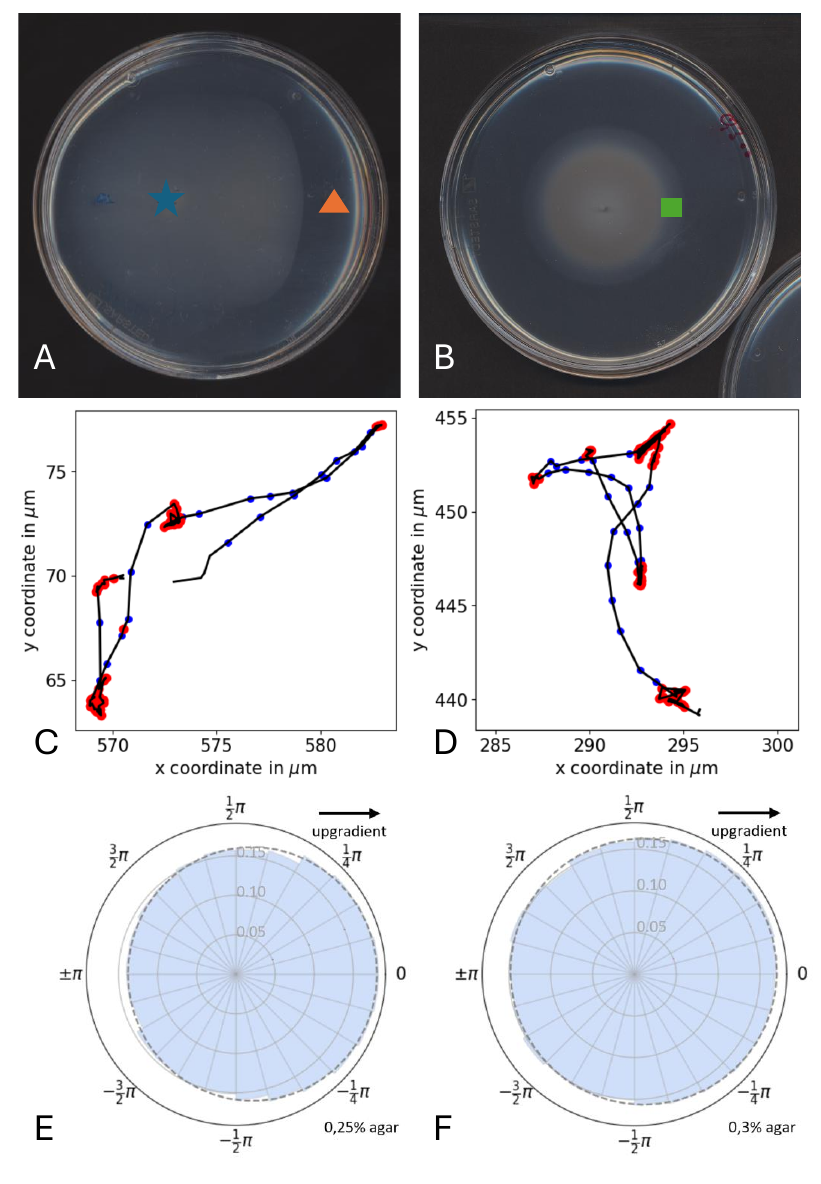}
\vspace{-0.4cm}
\caption{\textbf{Chemotaxis of \textit{P.~putida} on the population and the trajectory level.} Macroscopic spreading of a cell colony in $0.3 \, \%$ agar with (\textbf{A}) an externally induced chemoattractant gradient and (\textbf{B}) evenly distributed chemoattractant. In panel \textbf{A}, the spreading is not isotropic, because the bacteria move directionally towards the source of chemoattractant~(injection point:~orange triangle). Diameter of the plates:~$85 \, \mbox{mm}$. The bacteria consume the chemoattractant and create a spatial gradient. Panels in the middle row show exemplary bacteria trajectories in panels \textbf{C} $0.25\, \%$ and \textbf{D} $0.3\, \%$ agar, recorded at the edge of the expanding colony~(green rectangle in panel B). Run phases are marked in blue and events in red. Panels \textbf{E} and \textbf{F} represent probability densities of all swimming directions~$\phi$ of bacteria during run phases for $0.25\, \%$ and $0.3\, \%$ agar, respectively. Chemotaxis of bacteria is reflected at the trajectory level by biased motion towards the right~(E: $54.2 \pm 0.02 \, \%$; F: $52.4 \pm 0.02 \, \%)$. Error bars, obtained by bootstrapping~\cite{efron_introduction_1994}, represent~$2\sigma$ intervals. Dashed lines represent fits to the data obtained via Fourier expansion to first order, $p(\phi) \approx [1 + 2|f_1|\cos(\phi - \psi)] /(2\pi) $, where the first nontrivial mode~$f_1$ is another measure for the chemotactic bias~(E:~$|f_1| = 0.067$ and F:~$|f_1| = 0.036$). }
\label{fig:macroscopic_chemotaxis}
\end{figure}

\normalsize
\normalfont 
\noindent
\textbf{\textit{P.~putida} performs chemotaxis in semisolid agar.}
We first performed a macroscopic spreading experiment to examine whether \textit{P.~putida} performs chemotaxis in semisolid agar.
For this purpose, we injected casamino acids into the agar to create a gradient via diffusion.
Cells were afterwards injected some distance away from the chemoattractant source~(see Methods for details).
Fig.~\ref{fig:macroscopic_chemotaxis}A shows that the colony did not spread isotropically.
Instead, an asymmetric spreading in the direction towards the source of chemoattractant was observed.
This clearly indicates that \textit{P.~putida} performs chemotaxis in the porous agar gel~\footnote{Note that this directional bias towards the attractant in this setting may be partly due to cell growth. At the back of the colony, less casamino acids are present leading to a decreased growth rate, which may affect the culture spreading~\cite{Wolfe1989a}. Thus, the preferred spreading direction may not only depend on motility but also on local differences in growth rate~\cite{Cremer2019}. However, a chemotactic drift was also observed at the level of single cell trajectories. As cell division does not play any role on the time scale at which individual cell trajectories were recorded, we conclude that there is no doubt about the existence chemotactic drift in our data.}. 

When chemoattractants were evenly distributed in the agar network, the bacterial colony spreads isotropically, see Fig.~\ref{fig:macroscopic_chemotaxis}B.
The spreading speed depends on the size of the pores as well as the motility and the growth rate of bacteria~\cite{Wolfe1989a,Licata2016,Cremer2019, Pfeifer2022}.
In this experimental setup, a gradient of chemoattractant was created at the edge of the colony, as the casamino acids were consumed by the bacteria~\cite{Wolfe1989a}.

We furthermore recorded single cell trajectories of \textit{P.~putida} in agar and analyzed them based on the classification described in Ref.~\cite{datta2024intermittentrunmotilitybacteria}. 
Exemplary trajectories are shown in Fig.~\ref{fig:macroscopic_chemotaxis}C and~D. 
We analyzed microscopy data from the edge of a spreading colony~(green rectangle in Fig.~\ref{fig:macroscopic_chemotaxis}B), in which bacteria create their own chemoattractant gradient by consumption of casamino acids. 
The angular histograms in Figs.~\ref{fig:macroscopic_chemotaxis}E and~\ref{fig:macroscopic_chemotaxis}F show that bacteria were more likely to swim upgradient than downgradient during run phases. 
Thus, also on the level of single cell trajectories, we confirmed that \textit{P.~putida} performed chemotaxis in both tested concentrations.
For the lower agar concentration of~$0.25 \, \%$, the bias was more pronounced than for~$0.3 \,\%$. 
Similar results were observed for the setup shown in Fig.~\ref{fig:macroscopic_chemotaxis}A.

\begin{figure*}[t]
\centering
\includegraphics[width=\textwidth, draft=false]{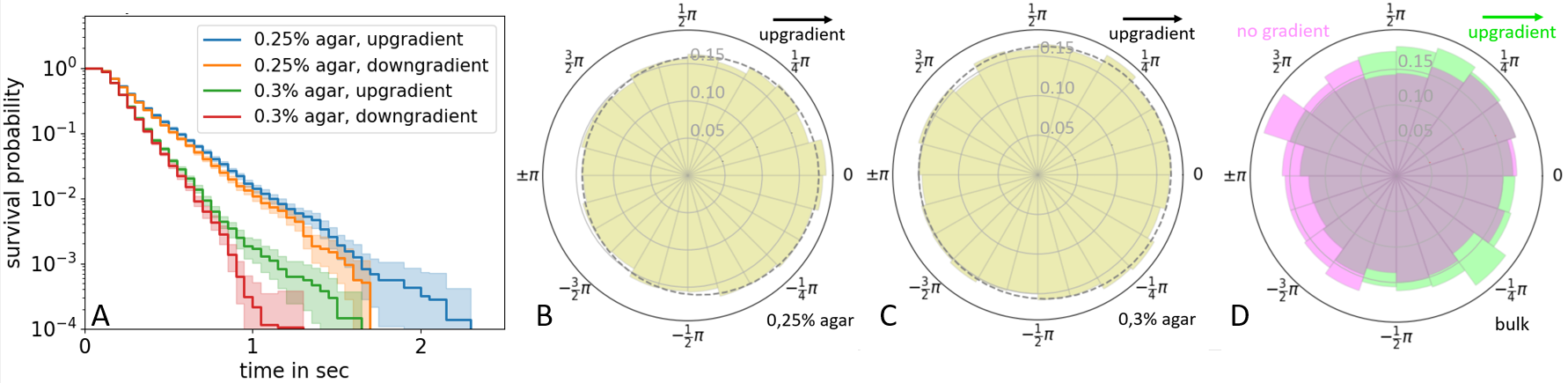}
\vspace{-0.6cm}
\caption{\textbf{Chemotaxis strategy of \textit{P.~putida}: run time and angle bias.} \textbf{A}: Run time distributions for different agar concentrations, represented as survival probability, i.e.~the probability that a run is longer than a given time, obtained by nonparametric maximum likelihood estimation~\cite{Vardi1982}. Error bars show a~$2\sigma$ interval, calculated by bootstrapping~\cite{efron_introduction_1994}. Panels~\textbf{(B)} and~\textbf{(C)} represent the probability density of the direction of runs~(orientation of the displacement vector from the first to the last data point of each run). \textbf{B}:~$53.3 \pm 0.3\%$ of 68431 runs~(from 17703 tracks) in~$0.25\%$ agar are oriented upgradient~(to the right in this case). The dashed line represents a fit to the data obtained via Fourier transform to first order, $p(\phi) \approx [1 + 2|f_1| \cos(\phi - \psi)] / (2\pi)$, where the first nontrivial mode~$|f_1| = 0.054$ is a measure for the strength of the bias. In~$0.3\%$ agar~(panel \textbf{C}), $51.9 \pm 0.3\%$ of 73531 runs~(from 18714 tracks) are oriented upgradient; corresponding fit parameter:~$|f_1| = 0.029$~(dashed line). \textbf{D}: Probability density of the orientation of runs, recorded in bulk fluid~(data taken from Ref.~\cite{Alirezaeizanjani2020}). Green histogram:~orientation of runs in the presence of a gradient~(2450 runs, 1378 tracks); $54.8 \pm 1.1\%$ of runs are pointing in upgradient direction. The magenta histogram shows the control experiment without a gradient~(3363 runs, 1799 tracks); no significant bias is observed in the absence of a gradient~($49.4 \pm 1.0 \, \%$ of runs are oriented upgradient). }
\label{fig:microscopic_chemotaxis}
\end{figure*}

\vspace{3mm}
\noindent
\textbf{A run time bias is observed in agar, despite the limited mean free path length.}
The observation of chemotaxis, both at the population and trajectory level, raises the question which chemotaxis strategy is employed by \textit{P.~putida} to achieve this directed motion in agar. 
The gel network constitutes a disordered environment with an average pore size in the range of $740$nm to $4800$nm for $0.25\%$ agar~\cite{Licata2016}, comparable to the measured mean run length of bacteria $\ell \approx 8 \, \mu\mbox{m}$~\cite{datta2024intermittentrunmotilitybacteria}. 
Notably, the mean run length and time in agar are one order of magnitude smaller than in bulk liquid~\cite{Alirezaeizanjani2020} due to spatial confinement~(see Tab.~\ref{tab:mean_run_time}). 

To test for the presence of the well-known bacterial chemotaxis strategy via a run time bias despite the limited mean free path length, we analyzed the statistics of the duration of runs pointing ugradient and downgradient separately. 
In Fig.~\ref{fig:microscopic_chemotaxis}A, the inferred probability distributions are shown. 
The run time distributions approximately decay exponentially. 
The mean run time strongly depends on the agar concentration:~shorter mean free path length in the denser agar network results in a mean run time that is significantly shorter in~$0.3 \, \%$ agar than in~$0.25 \, \%$ agar, cf.~Tab.~\ref{tab:mean_run_time}. 
Comparing the statistics of run times for up- and downgradient runs nonetheless reveals a small run time bias for both agar concentrations; runs upgradient are indeed longer than downgradient runs on average. 

Based on the mean run times of upgradient runs~$\tau_{\text{up}}$, downgradient runs~$\tau_{\text{down}}$, and all runs in one agar concentration~$\tau_{\text{all}}$ as shown in Tab.~\ref{tab:mean_run_time}, the strength of the run time bias can be quantified as~$S = \left(\tau_{\text{up}} - \tau_{\text{down}} \right) / \tau_{\text{all}}$.
We find~$S_{0.25} = 0.037 \pm 0.004$ for~$0.25\%$ agar and~$S_{0.3} = 0.022 \pm 0.002$ for~$0.3\%$ agar. Thus, the strength of the run time bias decreases with increasing agar density.

\begin{table}[b]
\centering
\begin{tabular}{c||c|c|c}
	& $\tau_{\text{all}}$ & $\tau_{\text{up}}$ & $\tau_{\text{down}}$\\
	\hline
	\hline
	$\;\;$ $0.25\%$ agar $\;\;$ & $\;\;$ $0.299 \, \mbox{s}$ $\;\;$ & $\;\;$ $0.304\, \mbox{s}$ $\;\;$ & $\;\;$ $0.293\, \mbox{s}$ $\;\;$ \\
	\hline
	$0.3\%$ agar & $0.232 \, \mbox{s}$ & $0.234 \, \mbox{s}$ & $0.229 \, \mbox{s}$
\end{tabular}
\caption{\textbf{Mean run times for upgradient runs and downgradient runs as well as all runs combined.} The statistical uncertainty was estimated via bootstrapping~\cite{efron_introduction_1994} to be~$3 \cdot 10^{-3} \, \mbox{s}$, corresponding to a~$2\sigma$ confidence interval. }
\label{tab:mean_run_time}
\end{table}

\vspace{3mm}
\noindent
\textbf{Runs are more likely to point towards increasing chemoattractant concentration both in agar and bulk liquid.}
Despite the restricted mean free path length in the porous agar matrix, a run time bias was detected.
Is this the only chemotaxis strategy at work or is \textit{P.~putida} relying on additional mechanisms to enhance the directional bias?
To test for a potential turn angle bias, we computed histograms showing the number of runs depending on their direction with respect to the chemoattractant gradient, see Figs.~\ref{fig:microscopic_chemotaxis}B and~\ref{fig:microscopic_chemotaxis}C~\footnote{Note that Figs.~\ref{fig:macroscopic_chemotaxis}E and~\ref{fig:macroscopic_chemotaxis}F show the orientation of all individual displacements during runs from one frame to the next, whereas Figs.~\ref{fig:microscopic_chemotaxis}B and~\ref{fig:microscopic_chemotaxis}C represent the histograms of the mean run orientations, i.e.~the number of runs in a certain direction. Therefore, Fig.~\ref{fig:macroscopic_chemotaxis} only indicates whether or not chemotaxis is present, but it does not reveal the underlying chemotaxis strategy---the bias may be due to run time or turn angle bias. In contrast, Figs.~\ref{fig:microscopic_chemotaxis}B and~\ref{fig:microscopic_chemotaxis}C enable us to conclude that there are more upgradient than downgradient runs, which indicates the presence of a turn angle bias}. 
The histograms show that, in addition to the run time bias, runs are more likely to point in upgradient direction than against it.
This suggests that \textit{P.~putida} may actively induce a turn angle bias, which increases the probability that a run is directed towards increasing chemoattractant concentration after a turn event.
The anisotropy in the histogram was observed for both agar concentrations, but it is pronounced for the less dense agar network.
Specifically, $53.3 \pm 0.3\%$ of runs were directed in upgradient direction for~$0.25\%$ agar, while this was only the case for~$51.9 \pm 0.3\%$ of runs for~$0.3\%$ agar.

\begin{figure*}[t]
\centering
\includegraphics[width=\textwidth, draft=false]{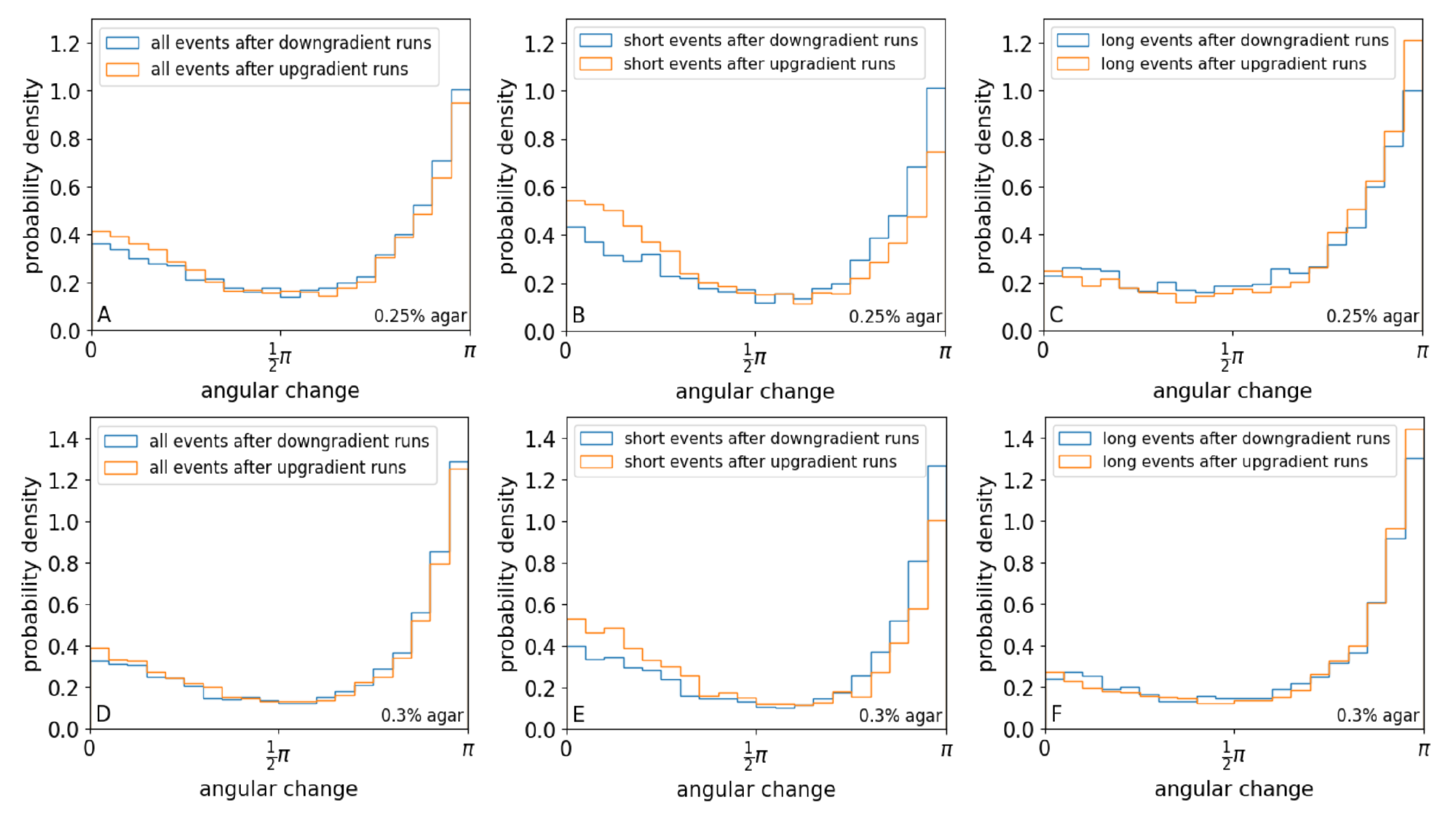}
\vspace{-0.8cm}
\caption{\textbf{Turn angle bias of \textit{P.~putida}.} The histograms show the angular reorientation of bacteria between two subsequent runs~(also referred to as an event) in dependence on the direction of the previous run in $0.25 \, \%$ agar~(top row) and in $0.3\,\%$ agar~(bottom row), respectively. Panels~\textbf{A} and~\textbf{D} include all observed events---only a small difference between initial upgradient~(orange lines) and downgradient~(blue lines) runs may be guessed. We further subdivided events according to their duration in short~(duration~$\leq 1\, \mbox{s}$; panels~\textbf{B} and \textbf{E}) and long events~(duration $> 1\, \mbox{s}$; panels \textbf{C} and \textbf{F}). While there is no turn angle bias for long events, the angular reorientation during a short event significantly depends on the initial run direction. Therefore, we conclude that short events, which are more likely to be actively triggered, motor induced turns, contribute a turn angle bias to the overall chemotaxis strategy of \textit{P.~putida}, whereas long events are dominated by spatial constrictions in the disordered agar gel and, thus, show no spatial dependence. } 
\label{fig:microscopic_mechanism}
\end{figure*}

We tested whether a similar bias can also be observed for chemotaxis in bulk fluid.
For this purpose, we reanalyzed the data from Ref.~\cite{Alirezaeizanjani2020} and found a similar anisotropy in the histogram of run directions, see Fig.~\ref{fig:microscopic_chemotaxis}D.
For the bulk fluid data, the results are less clear as the total number of observed runs is much lower.
Nevertheless, there is a clear difference in the number of runs pointing upgradient between the experiment with a nutrient gradient ($54.8\pm 1.1\%$) and the control experiment without a gradient ($49.4 \pm 1.0 \%$).
We may thus conclude that the bias in the direction of run episodes does not exclusively arise in a porous environment but is an intrinsic element of the chemotaxis machinery of \textit{P.~putida}.

\vspace{3mm}
\noindent
\textbf{A turn angle bias induces the orientational preference of runs in chemoattractant gradients.}
In order to achieve a directional bias, i.e.~a higher fraction of runs pointing towards the source of chemoattractant, cells have to actively tune their directional reorientation during a turn event depending on the direction of the preceding run phase (turn angle bias).
Indeed, our data indicated that if a bacterium swam towards increasing nutrient concentration (upgradient), the probability to observe only small turning angles during the subsequent turn event was slightly increased, thus increasing the chances that the bacterium continued to swim towards the nutrient source.
On the other hand, runs in downgradient direction were more often followed by turn events with a large change in direction, increasing the chances that the bacterium would turn towards the nutrient source, see Figs.~\ref{fig:microscopic_mechanism}A and~\ref{fig:microscopic_mechanism}D for $0.25 \,\%$ and $0.3 \,\%$ agar, respectively.
We therefore conjecture that {\it P.~putida} may actively modulate its turning angles, at least for part of its turn events.

\begin{figure}[t]
\centering
\includegraphics[width=\columnwidth, draft=false]{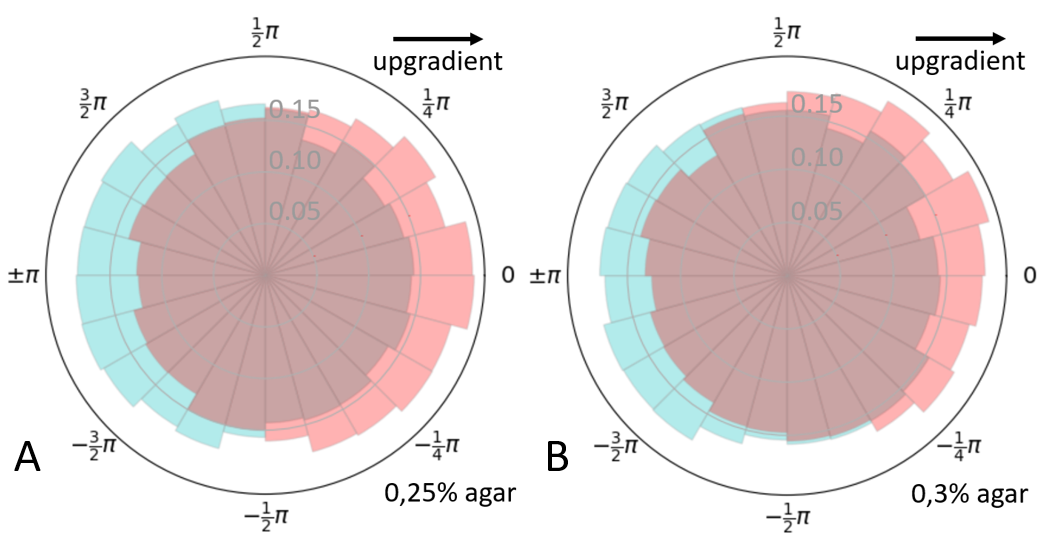}
\vspace{-0.5cm}
\caption{\textbf{Run orientation in dependence on preceding event duration.} The histograms show the probability density of run orientations, where runs are split into two subgroups:~runs following a short event~(duration~$\leq 1 \, \mbox{s}$) in red, runs following a long event~(duration~$> 1\, \mbox{s}$) in blue. The majority of runs following a long event~(trap) are oriented downgradient, i.e.~to the left. Panel~\textbf{A}: $0.25\, \%$ agar, panel~\textbf{B}:~$0.3\, \%$ agar. Fraction of runs, oriented upgradient, following a short event:~(A) $56.7 \pm 0.4 \, \%$ of $38376$ runs;~(B) $55.6 \pm 0.4 \, \%$ of $36351$ runs. Equivalent fractions of runs following a long event:~(A)~$44.9 \pm 0.5 \, \%$ of $17582$ runs;~(B) $46.8 \pm 0.4 \, \%$ of $25613$ runs. }
\label{fig:dependence_event_duration}
\end{figure}

\vspace{3mm}
\noindent
\textbf{Stops and reversals enhance the angle bias, while traps reduce it.}
It is known from previous studies of swimming motility in open bulk liquid that the runs of {\it P.~putida} are interrupted by stops and directional reversals (turn events with average turning angles of 0$^{\circ}$ and 180$^{\circ}$, respectively~\cite{Theves2013}).
The average duration of these actively triggered, motor-induced turning maneuvers is $0.3 \, \mbox{s}$ on average~\cite{Alirezaeizanjani2020}.
In the agar matrix, however, passive mechanical trapping is observed in addition, resulting in an average turn event duration of approximately~$2 \, \mbox{s}$~\cite{datta2024intermittentrunmotilitybacteria}.
To individually analyze the impact of motor-induced turning maneuvers and mechanical trapping on the angle bias, we separated the turning events according to their durations into short~(duration $<1 \, \mbox{s}$) and long~(duration $>1 \, \mbox{s}$) turn events and considered their turn angle distributions separately~(Fig.~\ref{fig:microscopic_mechanism}).
We found that the angle bias can be attributed exclusively to the short turn events in both agar concentrations~(see Figs.~\ref{fig:microscopic_mechanism}B and~\ref{fig:microscopic_mechanism}E), while the turn angle distribution does not depend on the direction of the preceding run for long events as shown in Figs.~\ref{fig:microscopic_mechanism}C and~\ref{fig:microscopic_mechanism}F.
This is supported by the results displayed in Fig.~\ref{fig:dependence_event_duration}:~runs after short turn events are more likely to point towards the nutrient source.
We thus conclude that actively triggered, motor-induced turns are required to induce an angle bias.

Note that, unlike short events, runs after long mechanical trapping events in the matrix are more likely to point away from the chemoattractant source~(Fig.~\ref{fig:dependence_event_duration}).
This can be explained by the typical trap geometry.
If a bacterium gets trapped, it is located in a narrow cavity that has, in many cases, only one exit through which the cell has also entered a trap.
As a result, many cells change their run orientation strongly after a trapping event because they have to reverse their swimming direction to leave the trap~\cite{datta2024intermittentrunmotilitybacteria}.
As overall more runs are pointing in gradient direction, the large angular change after a trapping event thus induces a tendency that the following run points away from the chemoattractant source.

\vspace{3mm}
\noindent
\textbf{Numerical simulations reveal importance of angle bias.}
In order to assess the importance of the two chemotaxis mechanisms---angle bias and run time bias---for the overall chemotactic drift, we performed agent based simulations of chemotactic spreading of a bacterial population.
Simulations are based on our recently proposed model of intermittent active motion~\cite{datta2024random} which we successfully applied to describe the diffusion of bacteria in a porous gel~\cite{datta2024intermittentrunmotilitybacteria}. 
Here, we extend this modeling framework to simulate the spreading of an ensemble of bacteria in a chemoattractant gradient. 
Specifically, we consider different conditions:~bacteria with both, run time and turn angle bias, are compared to simulated bacteria with turn angle bias only, run time bias only and, as a control, bacteria without any chemotactic response. 
Model parameters are estimated from experimental data, such as the experimentally observed run time, event duration and turn angle distributions.
For simulation details, we refer to the Supplementary Material~(SM)~\cite{noteSI}.

\begin{figure*}[t]
\centering
\includegraphics[width=0.8\textwidth]{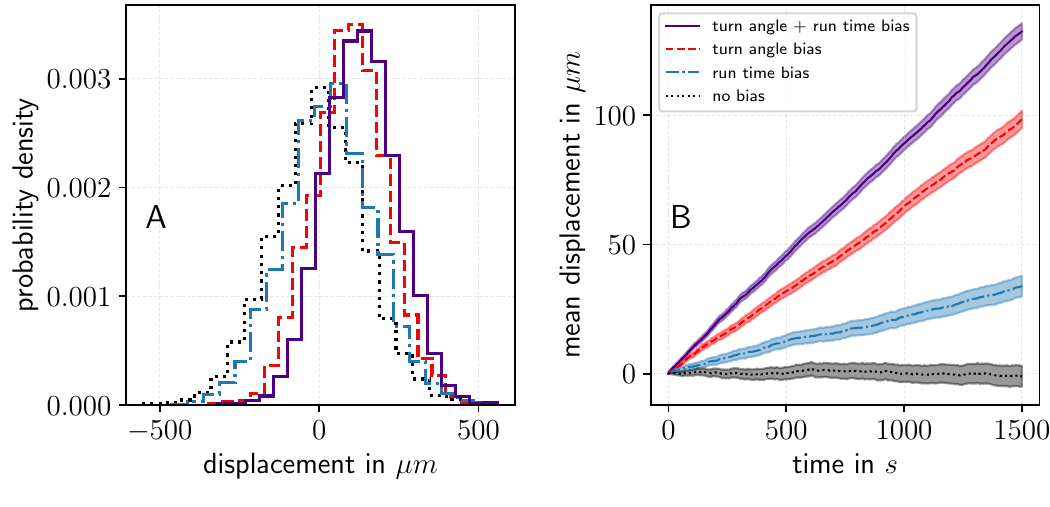}
\vspace{-0.4cm}
\caption{\textbf{Chemotactic drift of simulated bacteria.} Simulations were performed for bacteria without any chemotaxis strategy~(black), with run time bias only~(blue), with turn angle bias only~(red), and with both, run time and turn angle bias~(purple). \textbf{A}: Distribution of the total displacements of bacteria over the time of simulation~($25 \, \mbox{min}$). \textbf{B}: Mean displacement as a function of time, including a~$2\sigma$ standard error. } 
\label{fig:simulation}
\end{figure*}

To include run time and turn angle bias in the simulations, run times and angular changes during turn events were taken from the experimentally determined distributions in dependence on the preceding direction of runs with respect to the chemical gradient; results are displayed in Fig.~\ref{fig:simulation}.
We found that the average chemotactic drift of bacteria with turn angle bias only is approximately~$72 \, \%$ of the drift of bacteria relying on both strategies, whereas bacteria that rely on a run time bias only show a significantly slower chemotactic drift, reaching only~$25 \, \%$ of the drift of bacteria with both strategies.
We thus conclude that angle bias is the relevant mechanism that governs the chemotactic performance of bacteria in porous media such as agar gel.
Note that these simulations only provide conclusions regarding the efficiency of motility bias for chemotaxis and not about the colony growth in a swimming agar assay experiment, as the proliferation of bacteria is not included in the simulations~\cite{Wolfe1989a,Cremer2019}.

To compare the chemotaxis efficiency with the navigation strategy in bulk liquid, we also performed agent based simulations, considering the run time and turn angle statistics obtained previously~\cite{Alirezaeizanjani2020}.
We found that, although turn angle bias remains the predominant mechanism, it only contributes around~$59 \, \%$ of the total chemotactic drift velocity, whereas run time bias accounts for approximately~$41 \, \%$ of the drift of bacteria employing both strategies in bulk liquid~(cf.~SM~\cite{noteSI}). 
This underlines the importance of turn angle bias for chemotaxis in disordered environments, in which the run length is limit by the surrounding gel. \\

\noindent
\large{\textbf{DISCUSSION}}

\vspace{0.3em}

\normalsize
\normalfont 
\noindent
We have shown from the dynamics of a spreading culture as well as on the level of single cell trajectories that \textit{P.~putida} performs chemotaxis in a porous agar gel when exposed to gradients in casamino acid concentration.
We found that the chemotactic navigation relies on both run time bias and turn angle bias, where the latter is the dominant contribution to the overall chemotactic drift.

\vspace{3mm}
\noindent
\textbf{The role of turn angle bias in bacterial chemotaxis.}
The turn angle bias, as an additional chemotaxis strategy besides the run time bias, was previously observed for the peritrichously flagellated bacteria \textit{E.~coli}~\cite{Vladimirov2010, Saragosti2011, Sourjik2012, Pohl2017, Seyrich2018} and \textit{Salmonella} Typhimurium~\cite{Nakai2021} during chemotaxis in aqueous bulk fluid.
It has also been suggested as the dominating strategy for \textit{E.~coli} chemotaxis in porous media~\cite{Bhattacharjee2021}.
In this work, we have shown for the first time that a turn angle bias contributes to the chemotactic performance of a lophotrichously flagellated bacterial swimmer.
Thus, actively biasing the turning angle is not a unique feature of peritrichously flagellated bacteria but also plays a pivotal role during chemotaxis of polarly flagellated species.
In particular, our numerical simulations indicate that, in this crowded porous environment, the angle bias appears as the predominant chemotaxis strategy that accounts for almost $75 \, \%$ of the chemotactic performance.
Note that already a small adjustment in the angular change is sufficient to improve the chemotactic performance significantly~\cite{Vladimirov2010}.
The presence of turn angle bias may be crucially important to enable bacterial chemotaxis in disordered environments, in which the mean free path is reduced, in turn diminishing the influence of run time bias. 
Our findings are comparable to the results reported for \textit{E.~coli} chemotaxis in a porous medium, where the distribution of run directions suggests a similar mechanism, even though an angle bias was not directly measured~\cite{Bhattacharjee2021}.
In future work, more bacterial species should be examined for this phenomenon to substantiate whether chemotactic motion based on a turn angle bias is commonly observed in bacterial chemotaxis.

\vspace{3mm}
\noindent
\textbf{Flagellar dynamics and the mechanistic basis of turn angle bias.}
During run and tumble motility of \textit{E.~coli}, run phases of persistent forward motion are observed when all flagella are rotating counterclockwise.
Runs are interrupted by tumble events, when one or several individual flagella switch to clockwise rotation.
By this, the flagella bundle falls apart and bacteria change their body orientation.
The angular change during a tumble depends on the number of filaments that leave the flagellar bundle~\cite{Turner2000}:~the more flagella switch to clockwise rotation during a tumble, the higher, on average, the angular change.
To induce the chemotactic drift by tuning the tumble angle, the probability that a flagellum is rotating clockwise has to depend on the run direction with respect to the chemoattractant gradient.
If \textit{E.~coli} swims downgradient, the probability to switch to clockwise rotation is increased, so that the turning angle becomes larger.
When moving upgradient, less flagella switch to clockwise rotation, so that the angular change during a tumble will be smaller.
In this scenario, run time bias and turn angle bias are closely related because a higher probability to switch to clockwise rotation also leads to shorter runs in downgradient direction and vice versa~\cite{Vladimirov2010}. 
The question remains whether an angle bias can be also induced independent of a run time bias.

The motility pattern of \textit{P.~putida} is more complex and, therefore, the mechanism of turn angle bias proposed for \textit{E.~coli} cannot be directly transferred to \textit{P.~putida}.
\textit{P.~putida} swims in runs interrupted by stops and reversals.
During run phases, the flagellar bundle may rotate counterclockwise to push the cell body forward, it may rotate clockwise to pull the cell body, or it may wrap around it to induce a screw thread motion of the cell~\cite{Hintsche2017,Thormann2022}.
To induce transitions between these swimming modes, the flagellar motors have to switch between counterclockwise~(push) and clockwise rotation~(pull/wrapped) or they have to change from low~(pull) to high motor torque~(wrapped mode)~\cite{Hintsche2017,Park2022}.
We hypothesize that the turning angle during these transitions may depend on the degree of synchrony at which the motors are changing their torque or sense of rotation.
This is supported by recent numerical observations showing that a mismatch in motor torques may induce a rich variety of exotic swimming modes, some of which have been occasionally seen in experiments~\cite{Park2024}.
Moreover, also stop events may be triggered by asynchronous motor activity.
In particular for stops that interrupt run episodes in the wrapped mode, a broad turn angle distribution was observed~\cite{Alirezaeizanjani2020}.

Taken together, our results demonstrate that the lophotrichously flagellated soil bacterium \textit{P.~putida} primarily relies on a turn angle bias to navigate along chemical gradients in porous gel-like matrices.
While details of the underlying mechanism remain elusive, we assume that the degree of synchrony in motor operation during turning maneuvers provides the mechanistic basis of active turn angle regulation during chemotaxis of this polarly flagellated swimmer. \\

\noindent
\large{\textbf{METHODS}}

\vspace{0.3em}

\normalfont
\small 

\noindent
\textbf{Cell culture.}
For the experiments, the strain \textit{P.~putida} KT$2440_{\text{S267C}}$ FliC was used~\cite{Hintsche2017}. \textit{P.~putida} was grown over night in a shaking culture at~$30 \, ^\circ$C in LB medium.

\vspace{3mm}
\noindent
\textbf{Swimming agar assay with externally imposed gradient.}
To visualize chemotaxis macroscopically, a chemotaxis agar assay was used~\cite{Darias}. For this purpose,~$30 \, \mbox{ml}$ M8 ($6 \, \mbox{g}/\mbox{l}$ Na$_{2}$HPO$_{4}$, $3 \, \mbox{g}/\mbox{l}$ KH$_{2}$PO$_{4}$, $0.5 \, \mbox{g}/\mbox{l}$ NaCl) agar plates with $0.2 \, \%$ glucose and 1mM MgSO$_{4}$ as final concentration were poured into standard petri dishes ($\varnothing \, 85 \, \mbox{mm}$, height $15 \, \mbox{mm}$). Plates solidified over four hours. $10 \, \mu$l of $10 \, $ casamino acids were injected into the agar. The plates were kept overnight to create the chemoattractant gradient. After around $16$ hours, $2 \, \mu \mbox{l}$ of cells with an OD$_{600}$ of around 1 were injected into the agar with a distance of 0.5 to $3.5\, \mbox{cm}$ away from the casamino acids injection point. The effect of the nonisotropic spreading could be observed after one to two days. The effect could be seen for all tested distances between the chemoattractant and the culture injection point.

\vspace{3mm}
\noindent
\textbf{Swimming agar assay with self-generated gradient.}
To perform these experiments, we used the plate based assay from Ref.~\cite{Ha2014a}. M8 agar plates with $0.2 \, \%$ glucose, $0.5 \, \%$ casamino acids and $1 \, \mbox{mM}$ MgSO$_{4}$ as final concentration were poured into petri dishes. Plates solidified over four hours. $2 \, \mu \mbox{l}$ of cells with an OD$_{600}$ of around $1$ were injected into the agar. The agar plates were incubated over night at room temperature. The trajectories of bacteria were recorded at the edge of the spreading colony around 20 hours after injection of the bacteria.

\vspace{3mm}
\noindent
\textbf{Microscopy.}
An inverted microscope (Olympus IX71) was used for the microscopy recordings. The ORCA-Fusion BT Digital CMOS camera from Hamamatsu was used for recordings together with the Hokawo software (Hamamatsu). Phase contrast images were recorded $30 \, \mu \mbox{m}$ above the bottom surface with a 20x UPLFLN-PH objective (Olympus) at a framerate of $20$ frames per second. The individual recordings had a total length of 1 minute.

\vspace{3mm}
\noindent
\textbf{Cell tracking.}
The segmentation and cell tracking was realized by a custom-made computer code described in Ref.~\cite{Theves2013}, which is based on the algorithm presented in Ref.~\cite{CROCKER1996298}. Short trajectories (duration less than $1.5 \, \mbox{s}$) and trajectories with small total displacement (less than $1 \, \mu \mbox{m}$) were filtered out. For each agar concentration, 10 recordings with a duration of one minute from different plates and from 2 different days were used and analyzed. We recorded a total of 17703 trajectories for $0.25 \, \%$ agar with a median duration of 66 frames and 18714 trajectories for $0.3 \, \%$ agar with a median duration of 79 frames.

\vspace{3mm}
\noindent
\textbf{Event detection in agar.}
In Ref.~\cite{datta2024intermittentrunmotilitybacteria}, we described a method to differentiate run phases from stop, turn, and trap events:~k-means clustering in combination with a minimal distance of $2 \, \mu \mbox{m}$ between different events was used to distinguish runs and events. For the clustering, the change of direction of motion at a given data point over a shorter time interval (three frames), the change of direction of motion over a longer time interval (five frames) at that data point, and the speeds incoming and leaving that data point were used. Trajectories were not smoothed. 

\vspace{3mm}
\noindent
\textbf{Measurements and event detection in bulk fluid.}
The data and the event detection presented in Ref.~\cite{Alirezaeizanjani2020} were used for the analysis of chemotaxis in bulk liquid. Information on the methodology can be found there.

\vspace{3mm}
\noindent
\textbf{Agent based simulation.}
For agent based simulation, 5000 bacteria were simulated over a period of 30.000 time steps which is equivalent to~$25 \, \mbox{min}$ in the biological experiment. The experimentally measured run time distribution, turn angle distribution, event duration distribution, and the mean run speed were used to model the motility pattern of bacteria. The modeling framework was discussed in Ref.~\cite{datta2024random}; the data and inference techniques were also examined and shown in Ref.~\cite{datta2024intermittentrunmotilitybacteria}. 

We extended the modeling framework proposed in Ref.~\cite{datta2024random} to simulate chemotaxis as follows. For the simulations with angular bias, the turn angle distribution was divided into angular changes after upgradient runs and angular changes after downgradient runs; the corresponding random numbers were drawn from these distributions accordingly. For the run time bias, data were treated analogously: runs were divided into subgroups~(upgradient vs.~downgradient runs) and the run-time distributions were inferred and used separately. Further simulation details are discussed in the Supplementary Material~\cite{noteSI}. \\

\noindent
\large{\textbf{REFERENCES}}

\vspace{0.3em}

\vspace{1em}

\noindent
\large{\textbf{ACKNOWLEDGMENTS}}

\vspace{0.3em}

\normalfont 
\small
\noindent 
We acknowledge Kolja Klett and Hans Reimann for helpful discussions. This research was partially funded by Deutsche Forschungsgemeinschaft~(DFG), project ID 443369470-BE 3978/13-1~(VP and CB) as well as project ID 318763901~--~SFB1294~(AD, SB, RG and CB). \\

\noindent
\large{\textbf{AUTHOR CONTRIBUTIONS}}

\vspace{0.3em}

\normalfont 
\small
\noindent 
SB and VP conducted the experiments. SB and AD processed the experimental data and performed the analysis. AD performed the simulation. SB, VP, AD and CB wrote the manuscript. RG and CB supervised the project. All authors discussed the results and contributed to the final manuscript. \\

\noindent
\large{\textbf{COMPETING INTERESTS}}

\vspace{0.3em}

\noindent 
\small
\normalfont 
The authors declare no competing interests. \\ 

\noindent
\large{\textbf{DATA AVAILABILITY}}

\vspace{0.3em}

\noindent 
\small 
\normalfont 
The data that support the findings of this study are available from the corresponding author upon reasonable request. \\ 

\noindent
\large{\textbf{CODE AVAILABILITY}}

\vspace{0.3em}

\noindent 
\small 
\normalfont 
The computer codes used for simulations and numerical calculations are available from the corresponding author upon reasonable request. \\ 

\noindent
\large{\textbf{ADDITIONAL INFORMATION}}

\vspace{0.3em}

\small
\normalfont 
\noindent 
\textbf{Correspondence} and requests for materials should be addressed to CB. \\ 

\noindent
\large{\textbf{SUPPLEMENTARY MATERIALS}}

\vspace{0.3em}

\noindent 
\small 
\normalfont 
Supplementary material accompanies this paper. \\ 

\vspace{-0.2cm}

\noindent 
\textbf{Supplementary Note} (including Fig.~S1, Tabs.~S1-S2 and Refs.~\cite{datta2024random,datta2024intermittentrunmotilitybacteria,Alirezaeizanjani2020,Vardi1982}):~Agent based simulations \\ 

\end{document}